\documentclass{PoS}

\title{Double-scattering mechanism \\ 
of production of two $\rho^0$ mesons \\ 
in ultraperipheral, ultrarelativistic heavy ion collisions}

\ShortTitle{Double-scattering mechanism in UPC}

\author{\speaker{Antoni Szczurek}\thanks{Also at University of
    Rzesz\'ow, ul Rejtana 16, 35-959 Rzesz\'ow, Poland} and Mariola K{\l}usek-Gawenda\\
        H.Niewodnicza\'nski Institute of Nuclear Physics, Polish Academy
        of Sciences, ul. Radzikowskiego 152, 31-342 Krak\'ow, Poland \\
        E-mail: \email{Antoni.Szczurek@ifj.edu.pl}, \email{Mariola.Klusek@ifj.edu.pl}}

\abstract{We study, for the first time, differential distributions 
for two $\rho^0$ meson production in exclusive ultraperipheral, 
ultrarelativistic heavy ion collisions via a double-scattering mechanism. 
The calculations are done in the impact parameter space. 
The cross section for $\gamma A \to \rho^0 A$ is parametrized based on 
an existing model. Smearing of $\rho^0$ masses is taken into account. 
The results of calculations for single and double-$\rho^0$ production
are compared to experimental data at the RHIC and LHC energies. 
The mechanism considered gives a significant contribution 
to the $AA \to AA \pi^+\pi^-\pi^+\pi^-$ reaction. 
Some observables related to charged pions are presented too. 
We compare results of our calculations with the STAR collaboration results 
for four charged pion production.}

\FullConference{The European Physical Society Conference on High Energy Physics\\
		22--29 July 2015\\
		Vienna, Austria}

\def\Pom{{\bf I\!P}}
\begin{document}

\section{Introduction}

The exclusive production of simple final states
in ultraperipheral collisions (UPC) of heavy ions
is a special class of nuclear reactions \cite{reviews}.
At ultrarelativistic energies 
we can define two categories of the underlying reaction mechanisms.
First one is the photon-photon fusion \cite{rho0_gamma_fusion} 
and another one is double photoproduction of vector mesons \cite{DS_2014}.

\begin{figure}[htb]
\centerline{%
\includegraphics[scale=0.35]{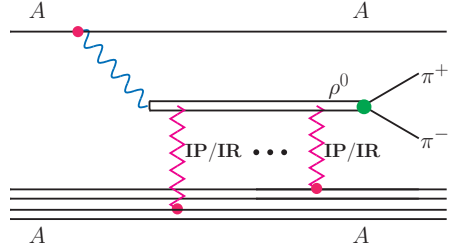}}
\caption{Single vector meson production by photon-Pomeron (or Pomeron-photon)
fusion.}
\label{Fig:gIP_rho0}
\end{figure}

In Fig. \ref{Fig:gIP_rho0} we show generic diagram of single nuclear 
$\rho^0$ production (and its decay into $\pi^+\pi^-$ state) 
via $\gamma$-Pomeron or Pomeron-$\gamma$ exchange mechanism
in UPC of heavy ion.
Photon emitted from a nucleus fluctuates into hadronic or quark-antiquark
components which rescatters in the second nucleus and converts into 
a simple final state.

\begin{figure}[htb]
\centerline{%
\includegraphics[scale=0.25]{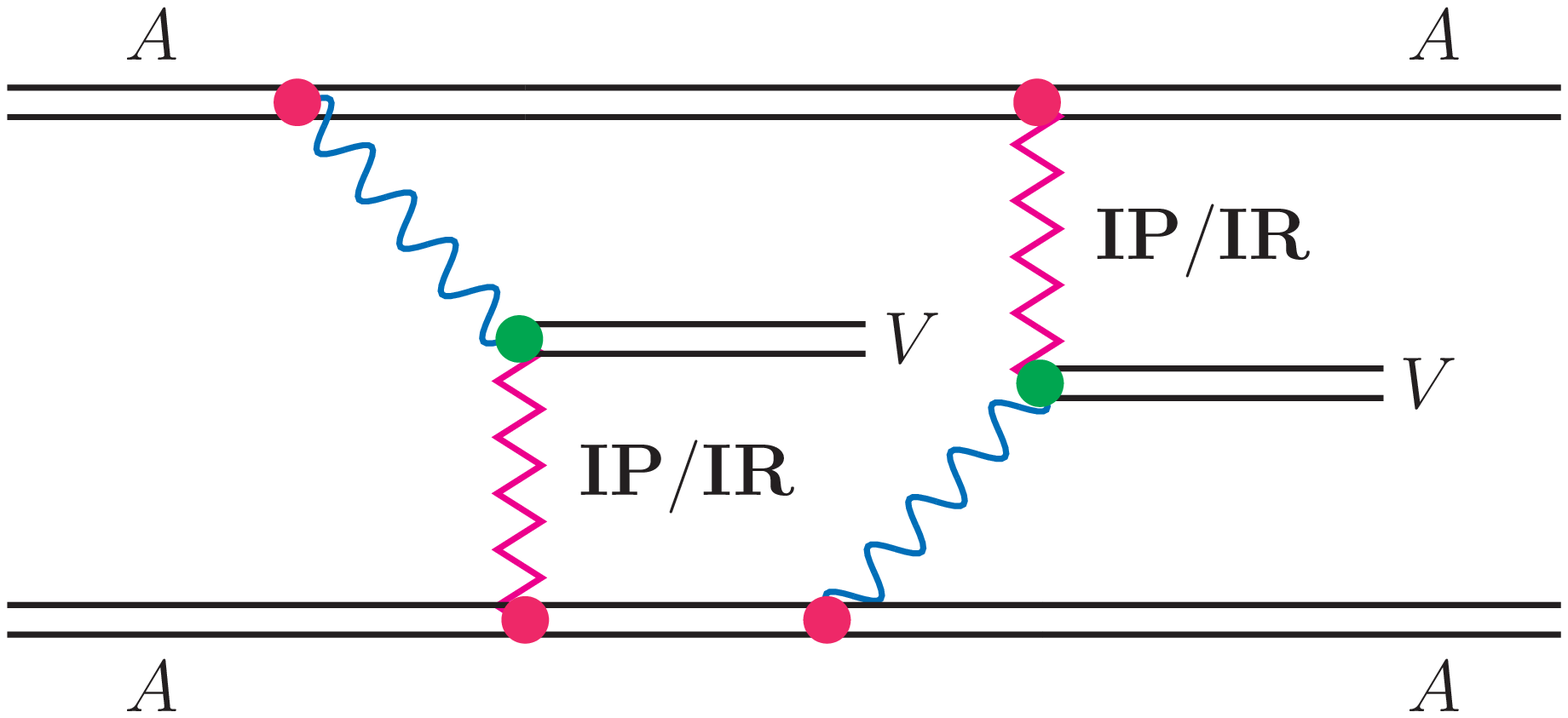}
\includegraphics[scale=0.25]{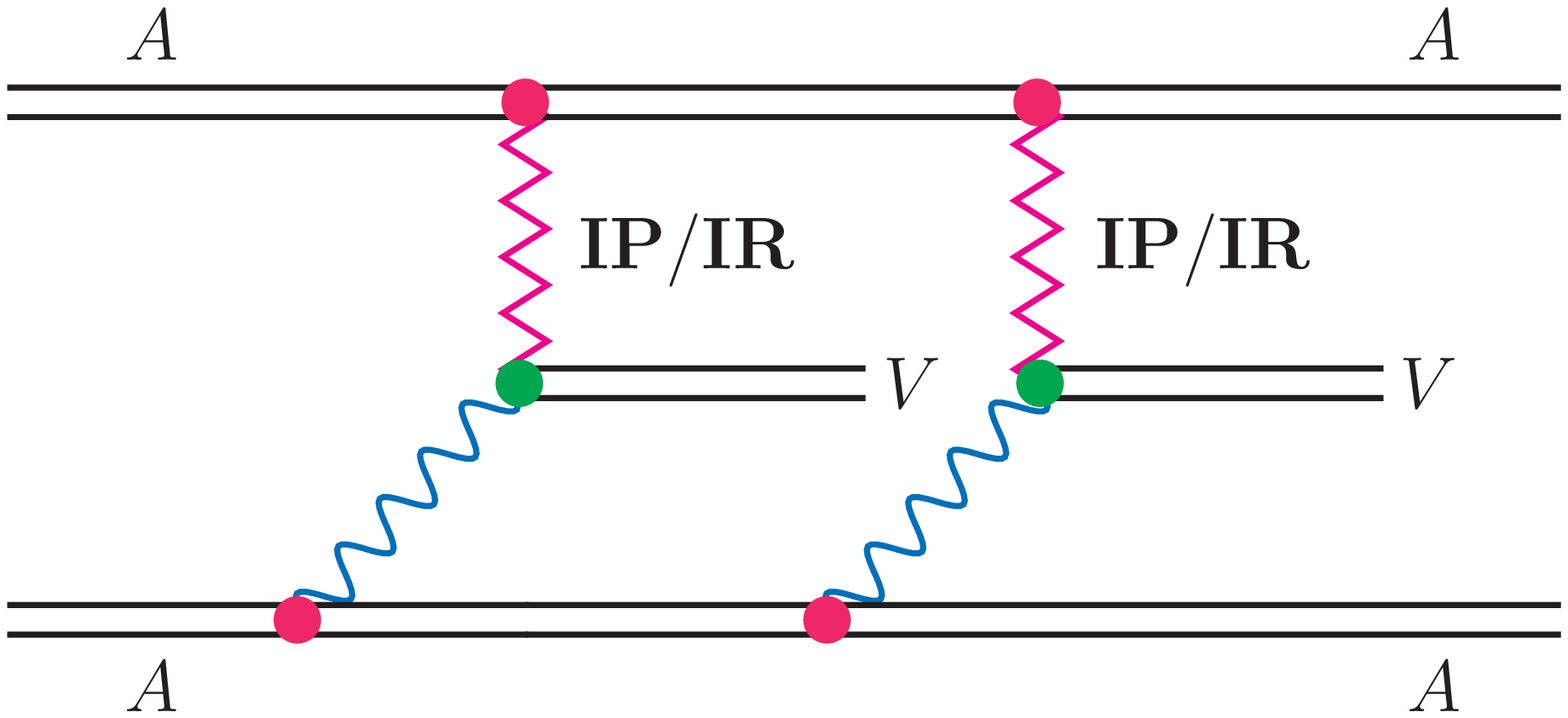}}
\caption{Double vector meson production by photon-Pomeron (or Pomeron-photon)
fusion.}
\label{Fig:gIP_rho0rho0}
\end{figure}

\begin{figure}[htb]
\centerline{%
\includegraphics[scale=0.35]{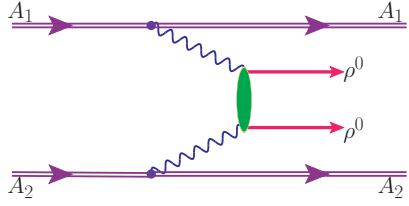}}
\caption{Double $\rho^0$ production in $\gamma\gamma$ fusion.}
\label{Fig:gg_rho0rho0}
\end{figure}

We study, for the first time, differential distributions 
for exclusive production of two $\rho^0(770)$ mesons 
(Fig. \ref{Fig:gIP_rho0rho0})
and its decay into four charged pions 
in double scattering (DS) processes. 
The results will be compared with the contribution 
of two-photon mechanism (Fig. \ref{Fig:gg_rho0rho0}). 
The analysis include a smearing of $\rho^0$ mass ($\Gamma \sim$ 0.15 GeV) 
using a parametrization of the ALICE Collaboration \cite{rho_smearing}.

\begin{figure}[htb]
\centerline{%
\includegraphics[scale=0.25]{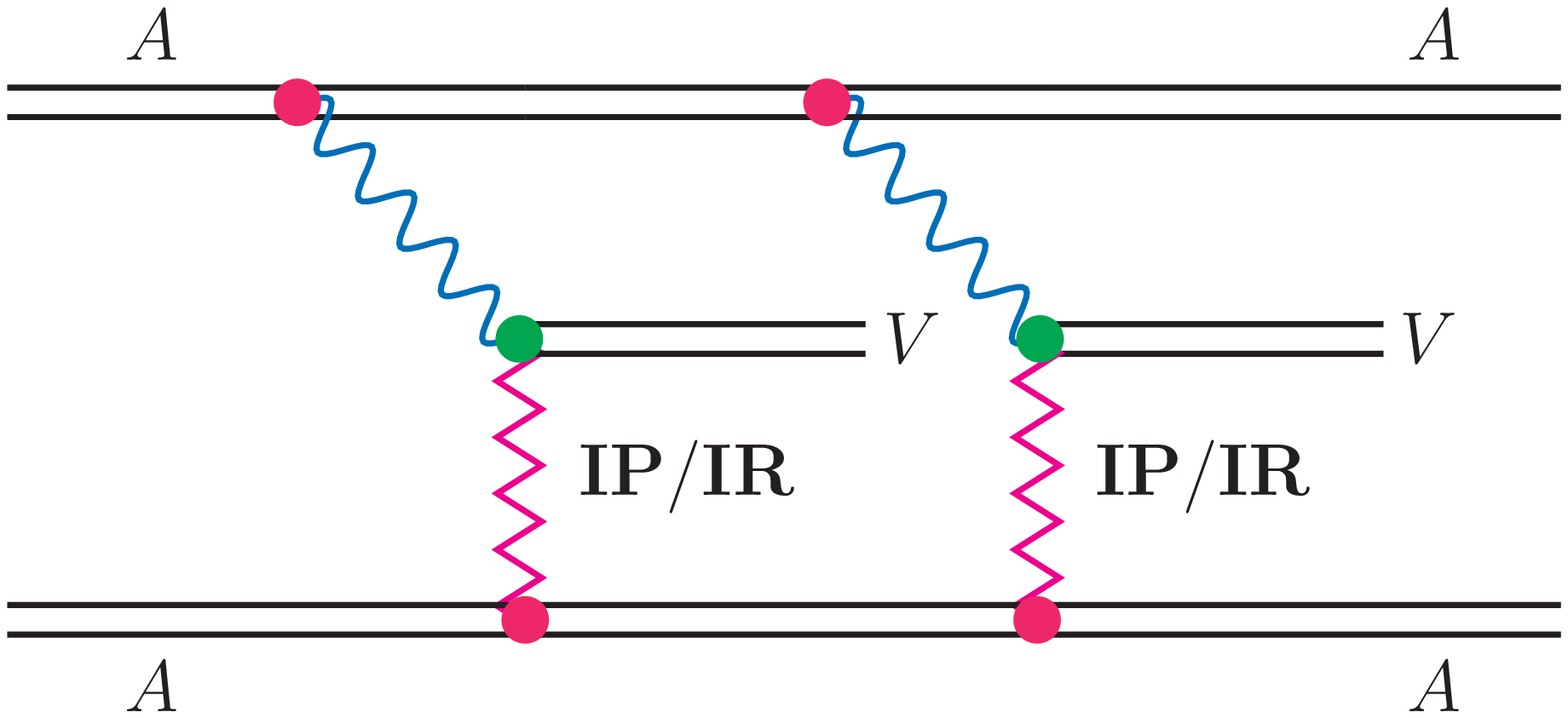}
\includegraphics[scale=0.25]{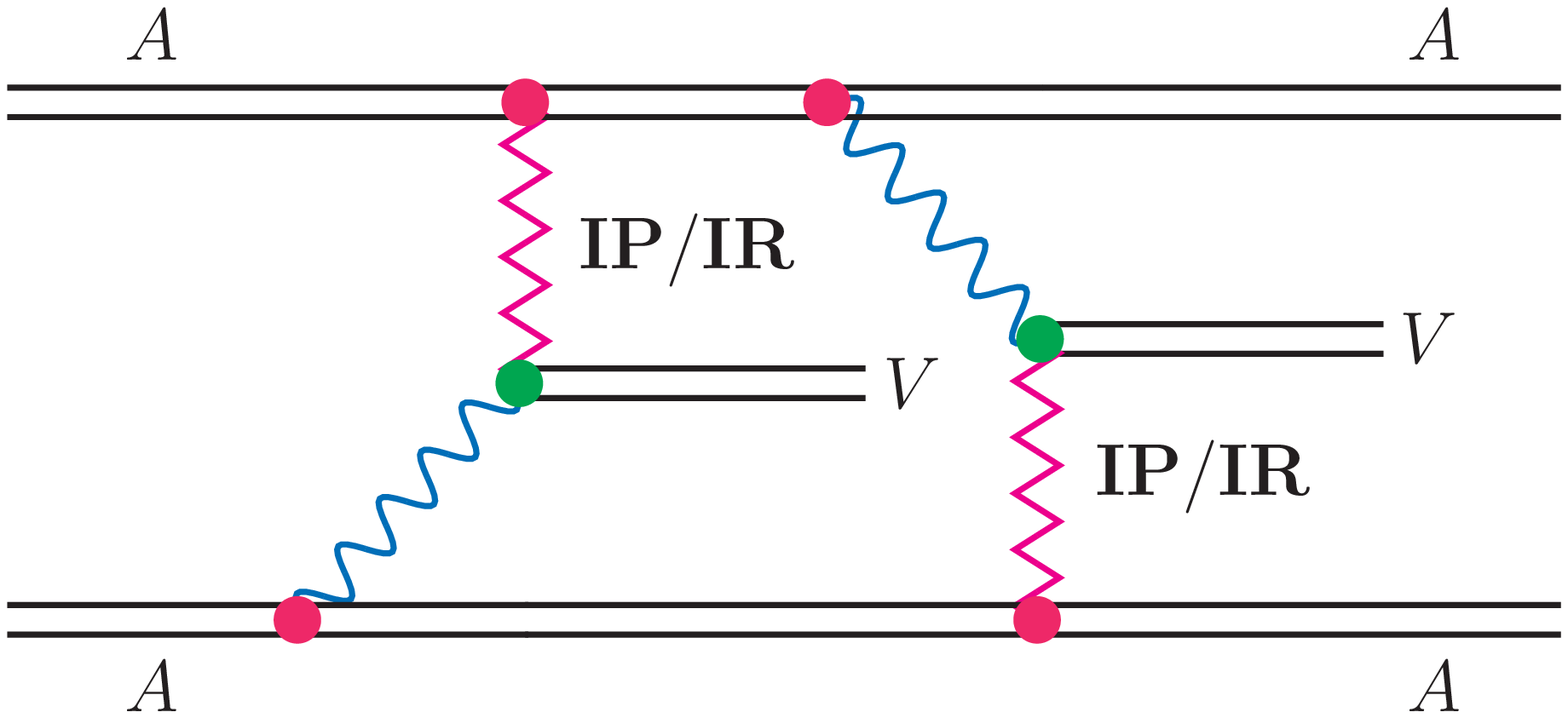}}
\end{figure}
\vspace{-1.cm}

\section{Formalism}

In our approach the cross section for single-vector-meson production 
(Fig. \ref{Fig:gIP_rho0}) 
depends on the impact parameter $b$ 
(distance between two colliding nuclei)
and on the $\rho^0$ meson rapidity $y$
\begin{equation}
\frac{\mathrm{d} \sigma_{AA \to AA \rho^0}}{\mathrm{d}^2b \mathrm{d}y}
= \frac{\mathrm{d} P_{\gamma \Pom} (b,y)}{\mathrm{d}y} 
+ \frac{\mathrm{d} P_{\Pom \gamma} (b,y)}{\mathrm{d}y} \;,
\label{eq:sigma_single}
\end{equation}
where $P_{\gamma \Pom / \Pom \gamma} (b,y)$ expresses the probability density
for producing a vector meson at rapidity $y$ for fixed impact parameter $b$
of the heavy ion collision. Each probability is the convolution
of a flux of equivalent photon and the $\gamma A \to \rho^0 A$ cross
section:
\begin{equation}
P_{\gamma \Pom / \gamma \Pom} (b,y) =
\omega_{1/2} \, N(\omega_{1/2},b) \, \sigma_{\gamma A_{2/1} \to \rho^0 A_{2/1}} \;.
\end{equation}
Photon flux $N(\omega_{1/2},b)$ depends on the energy of photon
and on impact parameter $b$ (heavy ion -- heavy ion distance).
Generally photon flux is expressed through nuclear form factor $F(q)$
which is related to charge distribution in the nucleus.
Details of different types of form factors and their application
into nuclear calculation one can be found 
in Refs. \cite{rho0_gamma_fusion,muons,Jpsi,MKG_thesis}.
To calculate the $\sigma_{\gamma A_{2/1} \to \rho^0 A_{2/1}}$
cross section we use a sequence of equations which are presented
in \cite{KN1999}. Constants for the underlying $\sigma_{\gamma p \to \rho^0 p}$ cross section are obtained from a fit to HERA data \cite{HERA}.
The $\sigma_{\rho^0 A}$ total cross section 
are calculated using classical mechanics formula
\begin{equation}
\sigma_{\rho^0 A} = \int \mathrm{d}^2 \textbf{r} 
\left( 1-\exp \left( -\sigma_{\rho^0 p} \, T_A\left( \textbf{r} \right) \right) \right)
\label{eq:CM}
\end{equation}
or quantum mechanical Glauber formula
\begin{equation}
\sigma_{\rho^0 A} = 2 \int \mathrm{d}^2 \textbf{r} 
\left( 1-\exp \left( - \frac{1}{2} \sigma_{\rho^0 p} \, T_A\left( \textbf{r} \right) \right) \right) \;,
\label{eq:QM}
\end{equation}
where $T_A\left( \textbf{r} \right)$ is nuclear thickness function 
and $r$ is distance between photon emitted from first/second nucleus 
and middle of second/first one. 


Having a formalism for the calculation of single-vector-meson
production, one can use this formula to calculate cross section
for double-scattering mechanisms of two-vector-meson production 
in ultrarelativistic, ultraperipheral collisions of heavy ions. 
The cross section for the double $\rho^0$ photoproduction
is expressed with the help of probability density
of single $\rho^0$ meson production
\begin{equation}
\frac{\mathrm{d} \sigma_{AA \to AA \rho^0 \rho^0}}{ \mathrm{d}y_1 \mathrm{d}y_2}
= \frac{1}{2}  \int\mathrm{d}^2b  \left[
\left( \frac{\mathrm{d} P_{\gamma \Pom} (b,y_1)}{\mathrm{d}y_1} 
+ \frac{\mathrm{d} P_{\Pom \gamma} (b,y_1)}{\mathrm{d}y_1} \right)
\times 
\left( \frac{\mathrm{d} P_{\gamma \Pom} (b,y_2)}{\mathrm{d}y_2} 
+ \frac{\mathrm{d} P_{\Pom \gamma} (b,y_2)}{\mathrm{d}y_2} \right) \right]  \;.
\label{eq:sigma_double}
\end{equation}
Here we take into account four combinations of $\gamma \Pom$ exchanges: 
$\gamma \Pom-\gamma \Pom$, $\gamma \Pom - \Pom \gamma$,
$\Pom \gamma - \Pom \gamma$ and $\Pom \gamma - \gamma \Pom$ 
(see Fig. \ref{Fig:gIP_rho0rho0}).

We are a first group which can calculate not only total cross section
for double-vector-meson production but also some differential distributions,
e.g. two-dimensional distributions in rapidities of both vector mesons
or in $\rho^0\rho^0$ invariant mass.
Knowing that the produced $\rho^0$ mesons decay, with almost $100\%$ 
probability, into charged pions, and including a smearing of 
the $\rho^0$ masses in our calculation,
we can compare our predictions with experimental data
for $AA \to AA \pi^+\pi^-\pi^+\pi^-$ processes.

\section{Results}

\begin{figure}[!h]
\centerline{
\includegraphics[scale=0.35]{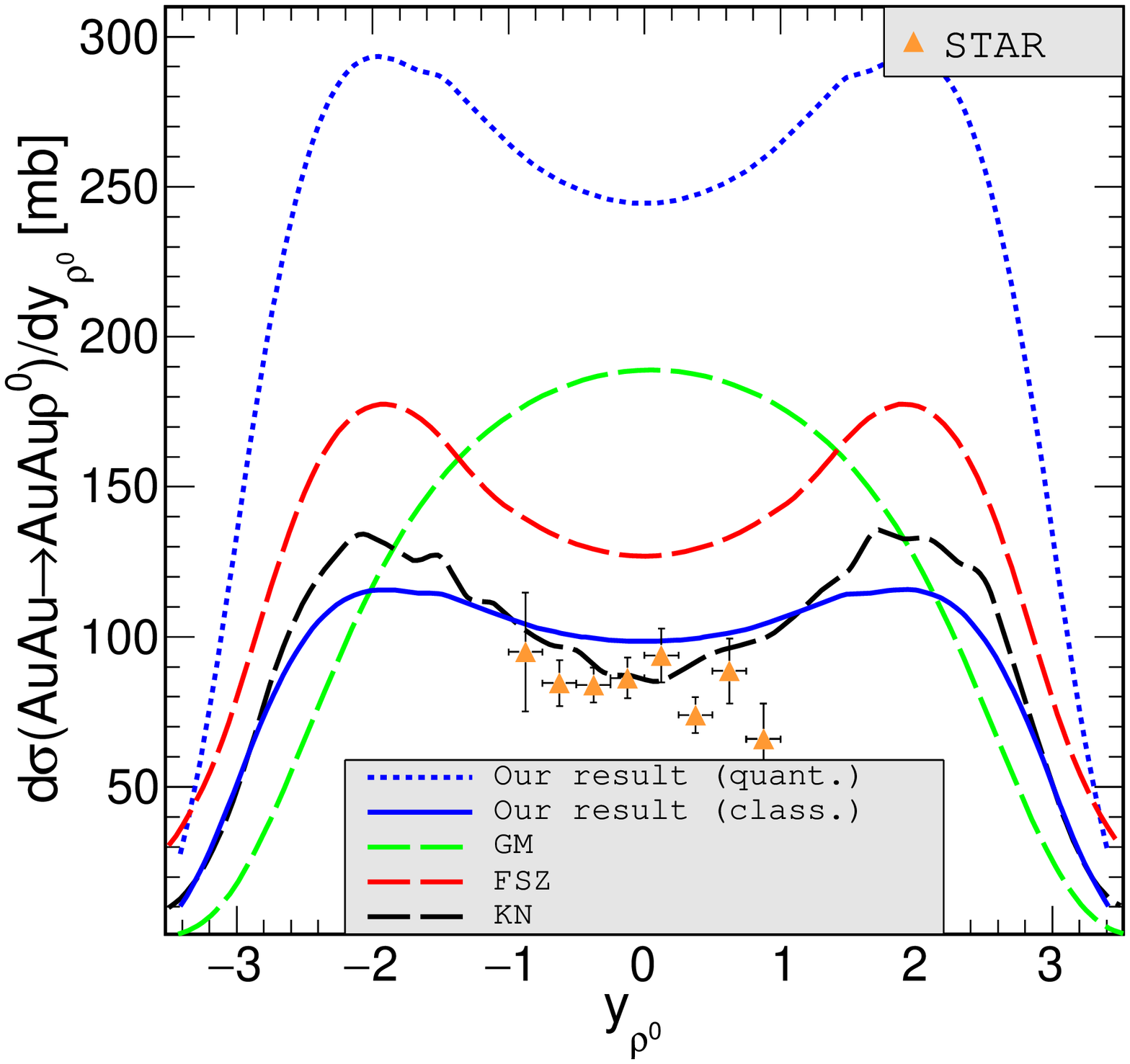}
\includegraphics[scale=0.35]{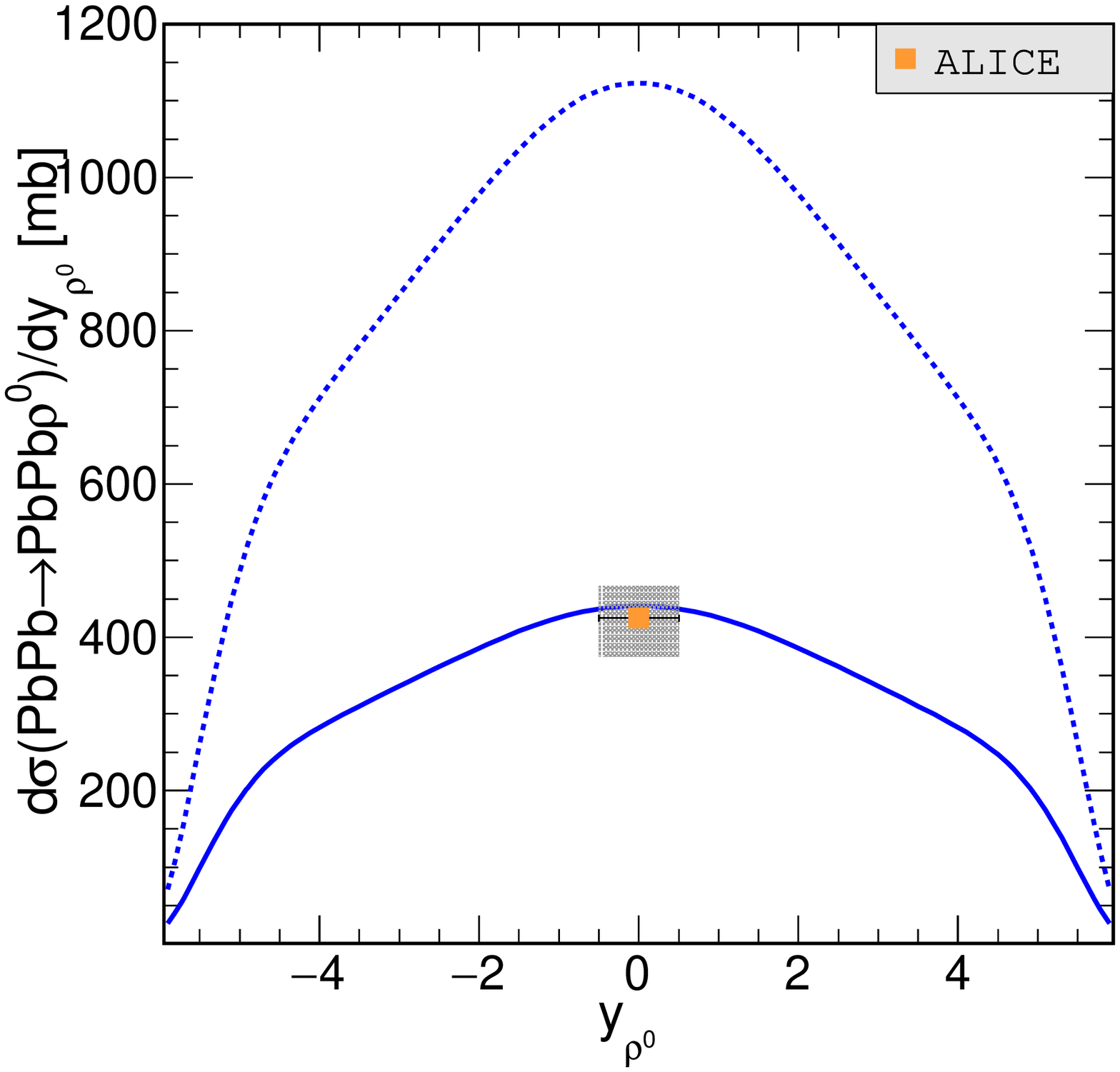}
}
\caption{Distribution in $\rho^0$ meson rapidity for single-$\rho^0$ production
for gold-gold collisions at RHIC energy (left panel) and
for lead-lead collisions at LHC energy (right panel).}
\label{Fig:dsig_dy_single}
\end{figure}

Fig. \ref{Fig:dsig_dy_single} shows a comparison of cross section
for coherent $\rho^0$ production measured by the STAR \cite{STAR} 
(left panel) and ALICE \cite{rho_smearing} (right panel) Collaborations
for different theoretical models (\cite{KN1999,GM, FSZ}).
One can observe that calculations for classical mechanics  
rescattering (Eq. \ref{eq:CM}) better (than in quantum approach 
(Eq. \ref{eq:QM})) 
describe both STAR and ALICE experimental data although we see no
deep theoretical reasons for this fact.
Our results (blue lines) relatively well describe the STAR and 
ALICE experimental data for the single vector meson photoproduction 
in ultrarelativistic heavy ion collisions (UPC).
This fact is very important for calculation of the cross section 
for double-scattering mechanism.

\begin{figure}[htb]
\centerline{
\includegraphics[scale=0.35]{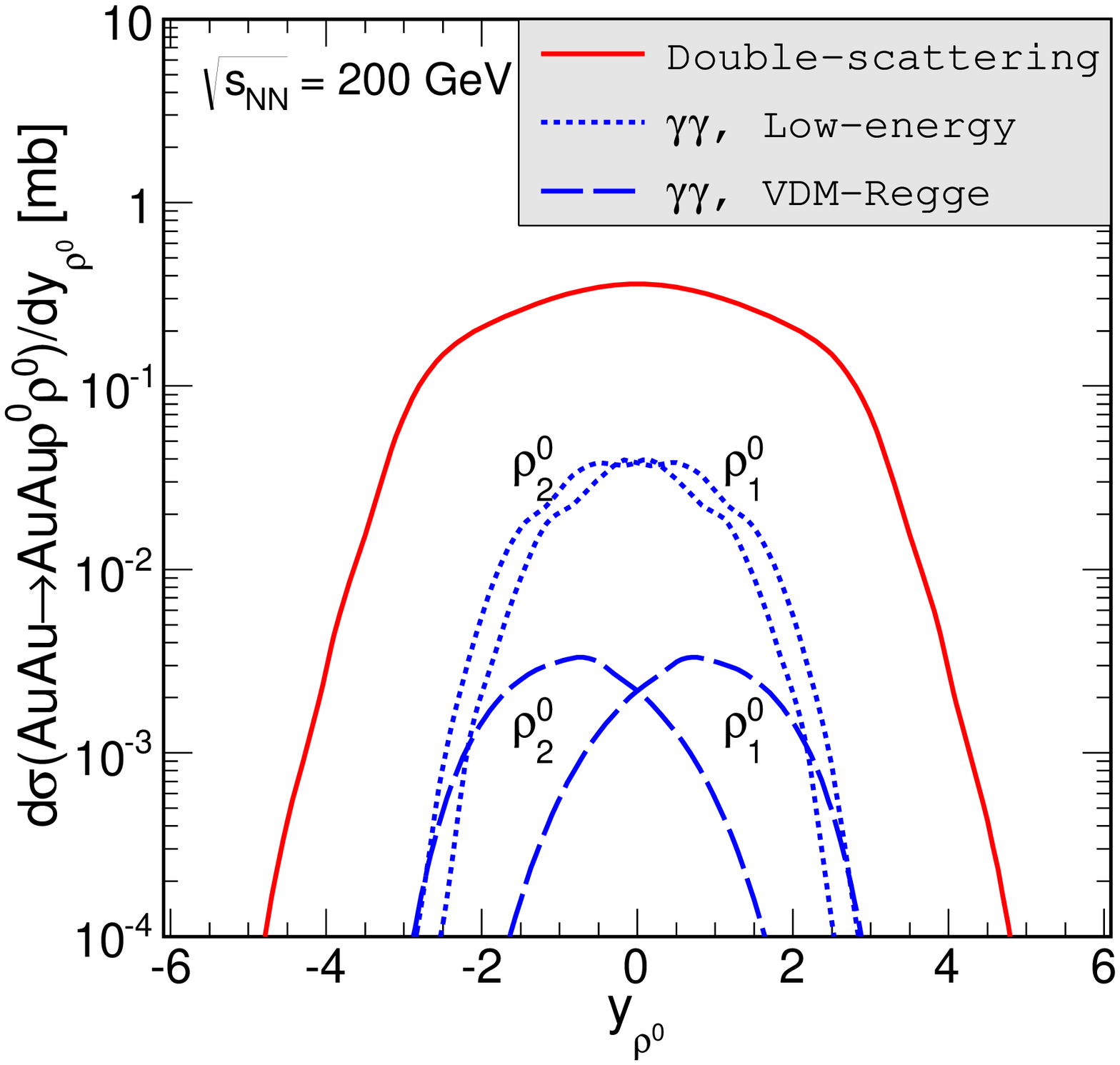}
\includegraphics[scale=0.35]{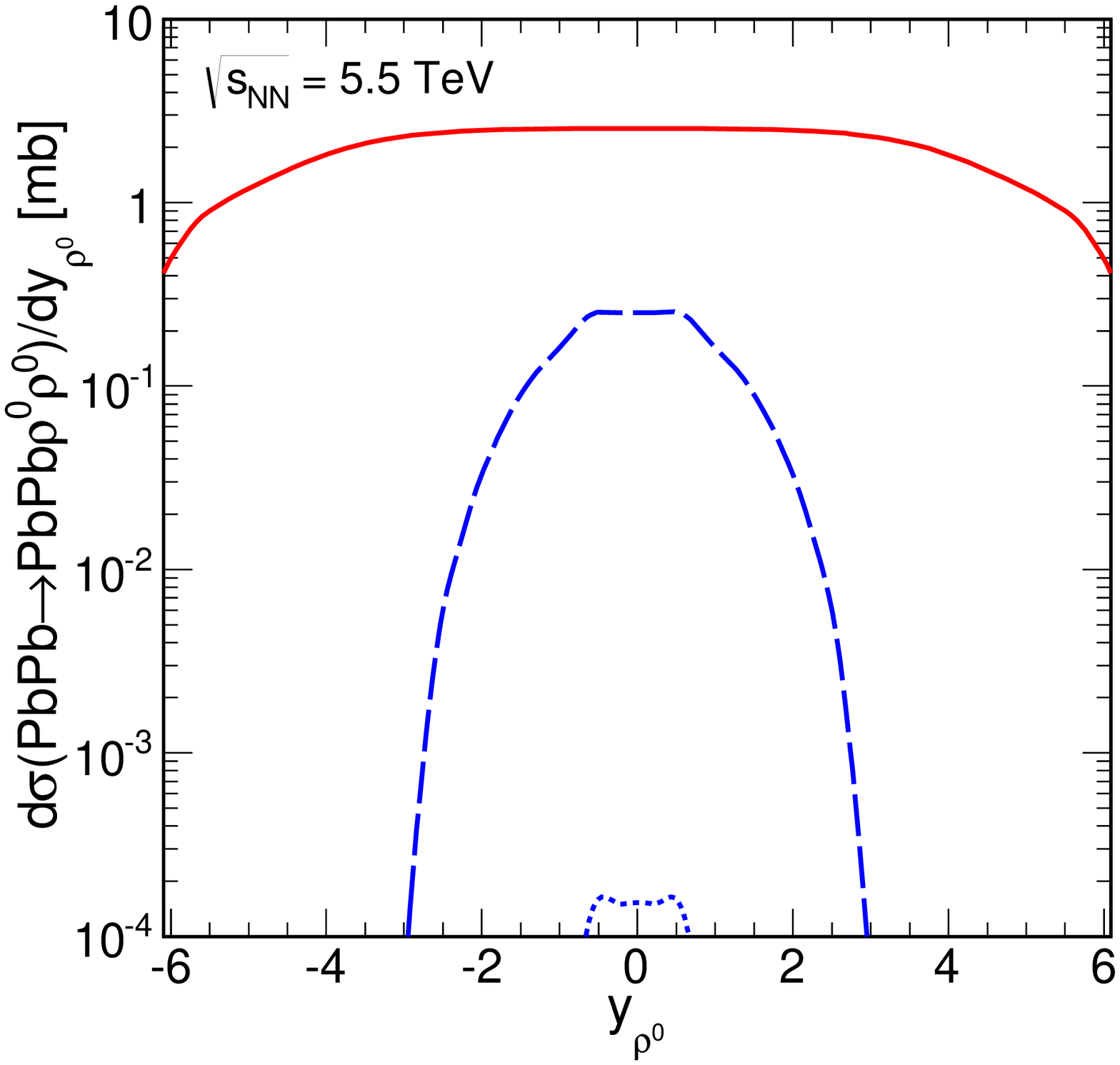}
}
\caption{Rapidity distribution of one of the $\rho^0$ meson produced 
in the double-scattering mechanism and in the $\gamma\gamma$ fusion at 
the RHIC (left panel) and at the LHC (right panel) energy.}
\label{Fig:dsig_dy_double}
\end{figure}

Fig. \ref{Fig:dsig_dy_double} shows differential cross section as
a function of one $\rho^0$ meson and a comparison of the 
results of the double-scattering and the $\gamma\gamma$ fusion mechanisms 
at RHIC (left panel) and at LHC (right panel) energy. 
The cross section for exclusive $\rho^0\rho^0$ production in
the $\gamma\gamma$ fusion approach is closely explained 
in Ref. \cite{rho0_gamma_fusion}.
There the elementary cross section ($\gamma \gamma \to \rho^0\rho^0$) 
is divided into two parts: low-energy component
($W_{\gamma\gamma}=(1-2)$ GeV) of not completely understood origin
and the VDM-Regge parametrization ($W_{\gamma\gamma}>2$ GeV) 
\cite{rho0_gamma_fusion}. 
In Fig. \ref{Fig:dsig_dy_double} one can observe a clear dominance of 
the DS component over the $\gamma\gamma$ component. 
The distribution for the center of mass energy $\sqrt{s_{NN}}=5.5$ TeV 
is much broader than that for $\sqrt{s_{NN}}=200$ GeV.
At the LHC energy the higher values of two-meson invariant mass 
becomes more important what corresponds to larger values of particle
rapidity. Therefore the high-energy component of the elementary cross section 
dominates at the LHC energy. Somewhat surprising at this energy is 
the fact that the VDM-Regge component is about three orders of 
magnitude larger than the low-energy component which is opposite to 
the case of the RHIC energy. 
Both at the RHIC and LHC energy, the contributions coming from 
the double-scattering mechanism is one order of magnitude
larger than that from the $\gamma\gamma$ fusion. 

\begin{figure}[htb]
\centerline{
\includegraphics[scale=0.35]{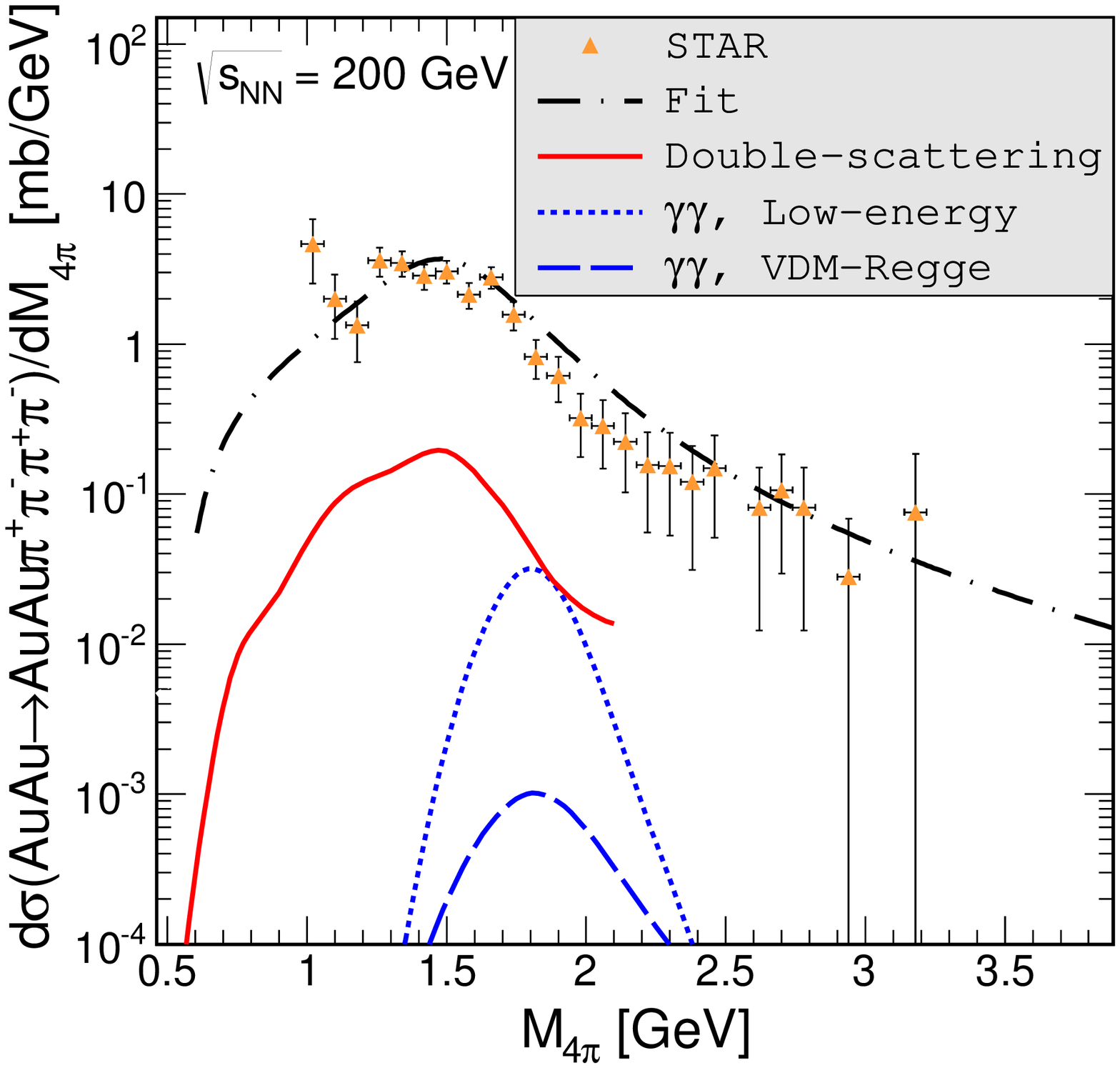}
\includegraphics[scale=0.35]{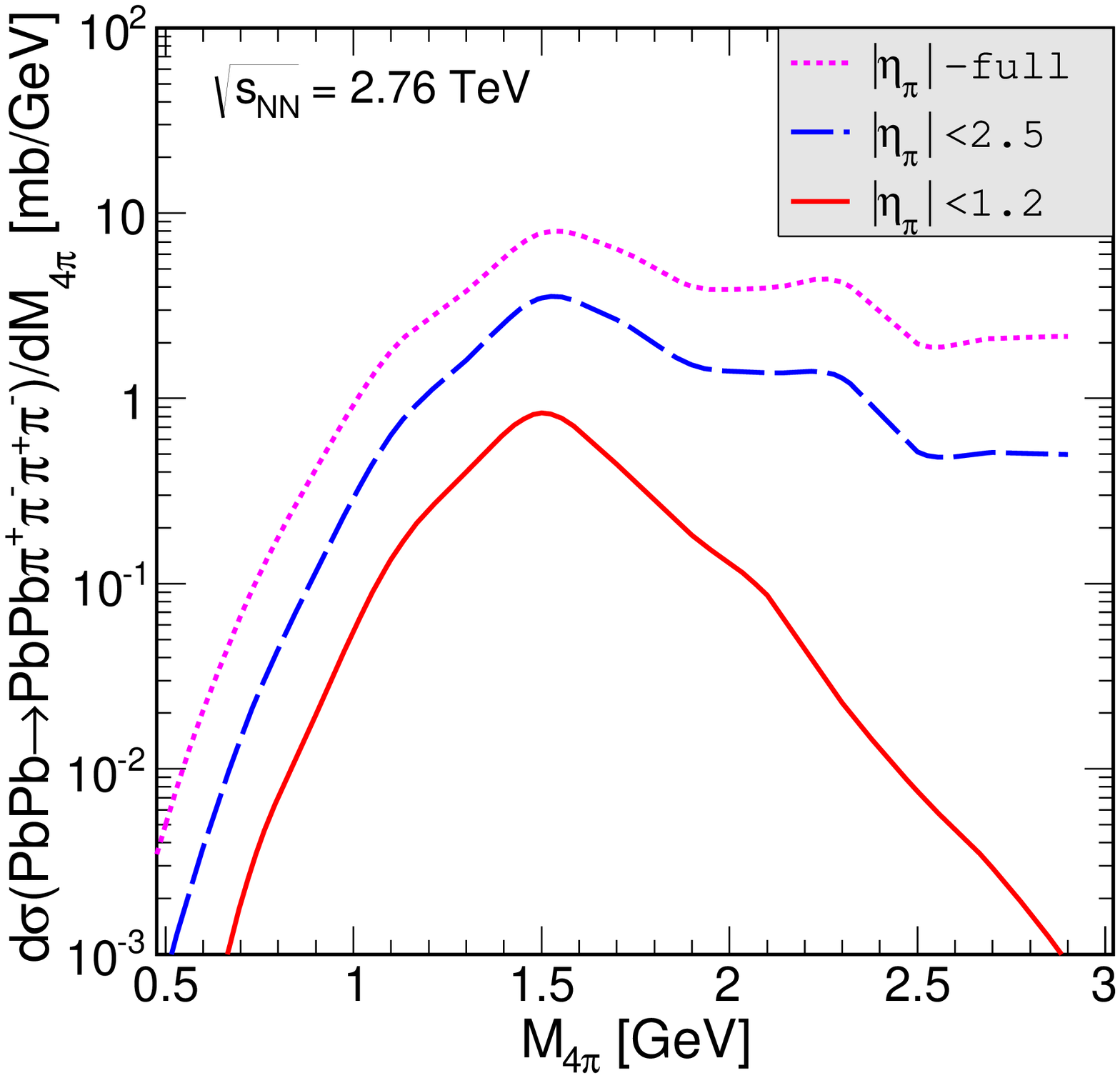}
}
\caption{Four-pion invariant mass distribution for the limited acceptance
of the STAR experiment (left panel) and for the limited range of 
pion pseudorapidity at the LHC energy (right panel).}
\label{Fig:dsig_dM4pi}
\end{figure}

The left panel of Fig. \ref{Fig:dsig_dM4pi} shows four-pion invariant mass 
distribution for double-scattering, low-energy bump and high-energy VDM-Regge 
$\gamma\gamma$ fusion mechanism for the limited acceptance of 
the STAR experiment ($|\eta_\pi| < 1$) \cite{STAR_4pi}. 
The double-scattering contribution accounts only for $20\%$ of 
the cross section measured by the STAR Collaboration. 
The dash-dotted line represents a fit of the STAR Collaboration. 
Probably the production of the $\rho^0(1450)$ and $\rho^0(1700)$ resonances
and their subsequent decay into the four-pion final state is 
the dominant effect for the limited STAR acceptance. 
Both, the production mechanism of $\rho^0(1450)$ and
$\rho^0(1700)$ and their decay into four charged pions are not yet understood. 
A model for production of the resonances and their decay has to be study
in the future.
The right panel of Fig. \ref{Fig:dsig_dM4pi} shows four-pion invariant mass 
distribution for double-scattering mechanism for the limited range of 
pion pseudorapidity at the LHC energy. The ALICE group collected
the data for four-charged-pion production with the limitation 
$|\eta_\pi|<1.2$, but we cannot compare our results with the ALICE data,
because those data are not absolutely normalized.

\section{Conclusions}

We have studied two-$\rho^0$ as well as four-pion production 
in exclusive ultrarelativistic heavy ion UPC,
concentrating on the double-scattering mechanism of 
single-$\rho^0$ production.
The produced two $\rho^0$ mesons give large contribution 
to exclusive production of the $\pi^+\pi^-\pi^+\pi^-$ final state.
We have compared contribution of four-pion production via
$\rho^0\rho^0$ production (double scattering and $\gamma\gamma$ fusion)
with experimental STAR data.
The theoretical predictions have similar shape 
as the distribution measured by the STAR Collaboration, but exhaust only
about $20\%$ of the measured cross section. 
The missing contribution can come from excited states of $\rho^0(770)$
and their decay into four charged pions.
We have discussed also a possibility of identifying 
the double scattering mechanism at the LHC.

In our calculation we have applied the smearing of $\rho^0$ meson 
masses by using the ALICE parametrization
\cite{rho_smearing} which is the most appropriate for the LHC data
(comparison of results for ZEUS, STAR and ALICE parameters
for relativistic Breit-Wigner and continuum amplitudes can be 
found in \cite{MKG_thesis}).



\end{document}